\documentclass[10pt,letterpaper]{article}
\usepackage[top=0.85in,left=2.75in,footskip=0.75in]{geometry}

\usepackage{amsmath,amssymb,gensymb}

\usepackage{changepage}
\usepackage{booktabs}
\usepackage{hyperref}
\usepackage{mathtools}
\newtagform{show_eq}{(Eq.\ }{)} 
\usetagform{show_eq}

\usepackage[utf8x]{inputenc}

\usepackage{textcomp,marvosym}

\usepackage{cite}

\usepackage{nameref,hyperref}

\usepackage[right]{lineno}

\usepackage{microtype}
\DisableLigatures[f]{encoding = *, family = * }

\usepackage[table]{xcolor}

\usepackage{array}
\pdfoutput=1
\usepackage[numbers]{natbib}

\newcolumntype{+}{!{\vrule width 2pt}}

\newlength\savedwidth

\raggedright
\usepackage[aboveskip=1pt,labelfont=bf,labelsep=period,justification=raggedright,singlelinecheck=off]{caption}

\makeatletter
\renewcommand{\@biblabel}[1]{\quad#1.}
\makeatother

\usepackage{lastpage,fancyhdr,graphicx}
\usepackage{epstopdf}
\fancyheadoffset[L]{2.25in}
\fancyfootoffset[L]{2.25in}
\usepackage{graphicx}
\usepackage{float}

\begin{document}

\vspace*{0.2in}

\begin{flushleft}
{\Large
\textbf\newline{Nanometer scale difference in myofilament lattice structure of muscle alter muscle function in a spatially explicit model} 
} 
\newline
\\
Travis Tune\textsuperscript{1*},
Simon Sponberg\textsuperscript{1,2},
\\
\bigskip
\textbf{1} School of Physics, Georgia Institute of Technology, Atlanta, Georgia, USA
\\
\textbf{2} School of Biological Sciences, Georgia Institute of Technology, Atlanta, Georgia, USA
\\
\bigskip

* ttune3@uw.edu

\end{flushleft}

\section*{Abstract}

Crossbridge binding, state transitions, and force in active muscle is dependent on the radial spacing between the myosin-containing thick filament and the actin-containing thin filament in the filament lattice. This radial lattice spacing has been previously shown through spatially explicit modeling and experimental efforts to greatly affect quasi-static, isometric, force production in muscle. It has recently been suggested that this radial spacing might also be able to drive differences in mechanical function, or net work, under dynamic oscillations like those which occur in muscles \textit{in vivo}. However, previous spatially explicit models either had no radial spacing dependence, meaning the lattice spacing could not be investigated, or did include radial spacing dependence but could not reproduce \textit{in vivo} net work during dynamic oscillations and only investigated isometric contractions. Here we show the first spatially explicit model to include radial crossbridge dependence which can produce mechanical function similar to real muscle. Using this spatially explicit model of a half sarcomere, we show that when oscillated at strain amplitudes and frequencies like those in the hawk moth \textit{Manduca sexta}, mechanical function (net work) does depend on the lattice spacing. In addition, since the trajectory of lattice spacing changes during dynamic oscillation can vary from organism to organism, we can prescribe a trajectory of lattice spacing changes in the spatially explicit half sarcomere model and investigate the extent to which the time course of lattice spacing changes can affect mechanical function. We simulated a half sarcomere undergoing dynamic oscillations and prescribed the Poisson's ratio of the lattice to be either 0 (constant lattice spacing) or 0.5 (isovolumetric lattice spacing changes). We also simulated net work using lattice spacing data taken from \textit{Manduca sexta} which has a variable Poisson's ratio. Our simulation results indicate that the lattice spacing can change the mechanical function of muscle, and that in some cases a 1 nm difference can switch the net work of the half sarcomere model from positive (motor-like) to negative (brake-like).


\section*{Author summary}

The myosin motors which are responsible for force generation in muscle not only produce axial force, but also produce radial force which can deform the myofilament lattice. Previous spatially explicit models investigated how this radial force and lattice spacing might influence isometric force, but were not able to generate net work under dynamic, phasically activated oscillations like those in \textit{in vivo} muscle, known as work loops. Here we revise a previous spatially explicit model and use it to investigate how the structure of the lattice spacing can affect whole muscle mechanical function during simulated work loops.

\linenumbers

\nolinenumbers

\section{Introduction}

In muscle, force is generated by the collective action of billions of myosin motors all undergoing nanometer scale conformational changes. However, the mechanical work output of a whole muscle, which is often the physiologically relevant parameter for animal locomotion, happens at the centimeter scale \cite{Josephson:1985um}. The multiscaled interactions of force, strain, binding, and activation are challenging but potentially tractable because muscle is a highly ordered, hierarchical tissue \cite{Campbell_Ken_2009}. For example, the interactions between chains of sarcomeres can produce emergent history-dependent behavior such as residual force enhancement that single sarcomeres might not \cite{Campbell2011a, Campbell2011b}. While this multiscale interplay has led to perhaps a greater understanding of molecular to macroscopic function in muscle than in any other tissue. However, it is challenging to extend this mechanistic understanding from quasi-static regimes to the dynamic behavior that makes muscle so versatile during movement. Here, we show in a spatially explicit, half-sarcomere model how the nanometer scale lattice structure of muscle can affect whole muscle mechanical function under dynamic conditions relevant for locomotion.

Tissue-scale physiological properties of whole muscle arise from the underlying 3D structure and geometry of muscle sarcomeres and myofilament lattice. For example, whole muscle's force-length relationship was originally attributed to the amount of overlap between myosin-containing thick filaments and the actin-containing thin filaments at the micron scale \cite{Josephson:1999tu, Huxley:1966}. This led to the sliding filament theory and allowed many of muscle's small-scale structure-function relationships to be inferred from whole muscle properties. However, it was later observed that the radial spacing between the thick and thin filaments was not constant, but rather could change with sarcomere axial strain changes \cite{Squire_muscle_review}. Not only that, but crossbridges (myosin motors bound to actin) can generate radial forces of comparable strength to axial forces, which in turn can deform the lattice \cite{Cecchi:1990, Millman1998, Maughan_GOdt_1981}. The probability of crossbridge attachment depends, in part, on the spacing between the filaments. Therefore, the radial spacing and crossbridges form a coupled system where strain changes imposed on the whole muscle can affect the thick-thin filaments spacing, which affects crossbridge binding. Taken together this means that variation in the radial separation of thick and thin filaments can contribute between 20\% - 50\% of the change in force in the quasistatic force-length curve compared to the contribution of axial overlap from sliding filaments\cite{Williams:2015, Williams:2010}. Radial separation of the thick and thin filaments is an important determinant of muscle force. 

Because these previous modeling and experimental efforts indicated the force-length curve of muscle is significantly affected by this radial spacing, we wondered if this radial separation could significantly affect a whole muscle's mechanical function under dynamic conditions such as those experienced during cyclic locomotion. Such function is often characterized by a muscle workloop \cite{Josephson_1993_WL_review, sponberg_perturbing_2023}, the net mechanical work during periodic contractions. This can, in part, be investigated experimentally. Because thick and thin filaments are arranged in a highly ordered hexagonal crystal lattice, the thick-thin spacing can be measured with high time resolved x-ray diffraction while simultaneously measuring force and length of the intact muscle under physiologically relevant conditions \cite{Irving:2000, Dickinson:2000vg, Iwamoto2018, Irving_review_2022}. Previously, we used this approach \cite{Tune_2020} to explore the differences in two muscles in the cockroach \textit{Blaberus discoidalis}, which have very similar quasistatic properties yet very dissimilar work outputs \cite{Ahn:2006dy}. We found that the two muscles have very similar lattice spacing under quasistatic conditions. However, when they cyclically contract they have different magnitudes and timing of lattice spacing changes. The differences in the force during workloops in the two muscles correlated to lattice spacing changes. This suggested that the nanometer-scale lattice spacing of a muscle could potentially explain the macroscopic whole muscle function \cite{Tune_2020}.

However, it is experimentally hard to show that a lattice spacing change can by itself change the work output of a whole muscle. While chemicals like dextran can be used to increase lattice spacing osmotically, this usually requires removing the cellular membrane ("skinning") \cite{Williams:2015}. Skinning the muscle makes isolating the effect of lattice spacing on mechanical work of the intact muscle difficult because the sarcolemma provides a stabilizing radial force to the lattice \cite{Wang_fuchs_1995, Gordon_review_2000}. So to test the effect of lattice spacing on muscle mechanical work output independent of other changes, we turned to a spatially explicit three-dimensional model of a muscle half sarcomere \cite{Daniel_1998, Tanner2007, Williams:2015}. The fact that the model is spatially explicit means that the model can allow us to investigate how the spatial arrangement of crossbridges, the biophysics of crossbridge formation at the nanometer scale, and the mechanical interaciton of nearby myosin heads on one another all affect force and mechanical work at the sarcomere scale. These models were initially developed to investigate how cooperativity between myosin heads could enhance force production in muscle, which is why it was necessary to make them spatially explicit, as opposed to a mass action model like MyoSim \cite{Campbell_myosim_2014}. This spatially explicit model was later used to explore work production under periodic contractions, however, that model had no dependence on radial spacing, which meant the effect of lattice spacing could not be investigated \cite{Tanner_2008}. Our model is based on a later model that was used to show that force-length properties are dependent on the lattice spacing \cite{Williams:2010, Williams:2013bi}. While this more recent model was able to produce good quasi-static results, which was the goal of those studies, it was unable to produce physiological realistic of mechanical work during oscillations. 

Since the lattice spacing of muscle has been previously implicated in whole muscle mechanical function \cite{Tune_2020}, here we use this spatially explicit modeling approach to test if nanometer differences in lattice spacing alone could have potentially significant impacts on whole muscle mechanical work. We first have to adapt previous models to produce reasonable work loops, periodically activated stress-strain curve, in a physiologically accurate range. We ground the model by comparing it to twitch and tetanus force responses as well as mechanical work at different phases of activation. As with previous modeling efforts, we use the physiological data from the dorsal longitudinal flight muscle (DLM) of \textit{Manduca sexta} \cite{Tu_SubMax, Tu_cardiac_manduca, Tanner_2008}. We chose this muscle not only because the twitch, tetanus, and work vs. phase of activation have been well established, but also because very detailed, high-resolution temporal measurements of the actin-myosin spacing changes simultaneous with work measurements have been obtained through x-ray diffraction \cite{Malingen2020, George:2013kg, Cass_sarc_breathing}.

In some muscles, such as invertebrate asynchronous flight muscle, the lattice spacing is approximately constant with length change \cite{Irving_Maughan:2000}. However, in many muscles, the lattice spacing depends strongly on length, in some cases expanding under an isovolumetric constraint \cite{Bagni1994, Dickinson:2000vg, Malingen2020, Tune_2020, Cass_sarc_breathing}. So not only might a static lattice spacing affect work, but muscles with different dependencies on length could have different relationships between net work and lattice offset and trajectory. In the spatially explicit model, we can prescribe different patterns of lattice spacing and axial strain to guide our examination of how lattice spacing changes affect net work. We first investigate the net work for lattice spacing offsets which remain constant over the course of the work loop, then we show the relationship between work and lattice spacing where the lattice spacing is isovolumetric -- the myofilament lattice expands or contracts proportionally to the axial strain as defined by a fixed Poisson ratio of 0.5. We then prescribe lattice spacing changes based on what has been observed \textit{in vivo} from \textit{M sexta} via x-ray diffraction. Our goal is to calculate net work under simulations with different lattice spacing offsets and trajectories to test if the lattice spacing changes on the scale of a single nanometer can modulate mechanical work, consistent with what was observed in the two cockroach muscles.





\section{Materials and methods}
\subsection{Model overview}
Our basis for the model is taken from \cite{Williams:2010}, \cite{JPowers_2018_model} and \cite{Williams:2013bi}. Each time step in the model follows a sequence of steps that ultimately give a scalable estimate of axial force produced by the myofilament lattice. Starting at the initial spatial configuration of the model, each myosin head first undergoes thermal forcing by drawing energies from a Boltzmann distribution for each spring that comprises the myosin head, which is then used to update the position of the heads. Then binding probabilities for each myosin head are calculated for the new spatial configuration of the half sarcomere and a set of prescribed rate equations (see below). After transitions between the states have been performed, the nodes which make up the thick and thin filaments undergo a minimization procedure to find the equilibrium configuration of the half sarcomere. This loop of diffusion, stochastic transition, and then force balancing is repeated at each time step. 

Earlier versions of the 2sXB spatially explicit model (termed 2sXB) investigated isometric muscle's force-length dependence on actin-myosin spacing \cite{Williams:2013bi, Williams:2010}. Those models were able to capture muscle's quasi-static behavior and to show that the force-length relationship in muscle is in fact highly dependent on radial spacing changes of actin and myosin which are coupled to changes in sarcomere axial length \cite{Williams:2010}. This is what led us to use that model to investigate if the actin-myosin spacing could have a significant effect on net work of a sarcomere. 

The net (mass-specific) mechanical work of muscle is given by the area enclosed by a stress-strain curve in which the muscle is periodically activated, called the muscle's work loop \cite{Ahn2012, Josephson:1985um, sponberg_perturbing_2023}. In work loop experiments, typically the \textit{in vivo} strain amplitude, frequency, and pattern of activation for a given muscle during a given behavior are measured in an intact animal, allowing the same patterns to be input into an excised muscle, from which net work can be measured \cite{Ahn:2006dy}. After establishing the behavior of the muscle under conditions which mimic its \textit{in vivo} behavior, the parameters of the work loop can be adjusted to explore the properties of muscle \cite{Josephson_1993_WL_review}. For example, the phase of activation can be adjusted, yielding a phase sweep. The phase of activation is the point in the length cycle when activation occurs. While the \textit{in vivo} range of phase of activation might be limited, by expanding the range of activation in work loops we can drive the muscle into different force producing regimes to examine its function. 

\begin{figure}[H]
\begin{centering}
\includegraphics[width = 1\linewidth]{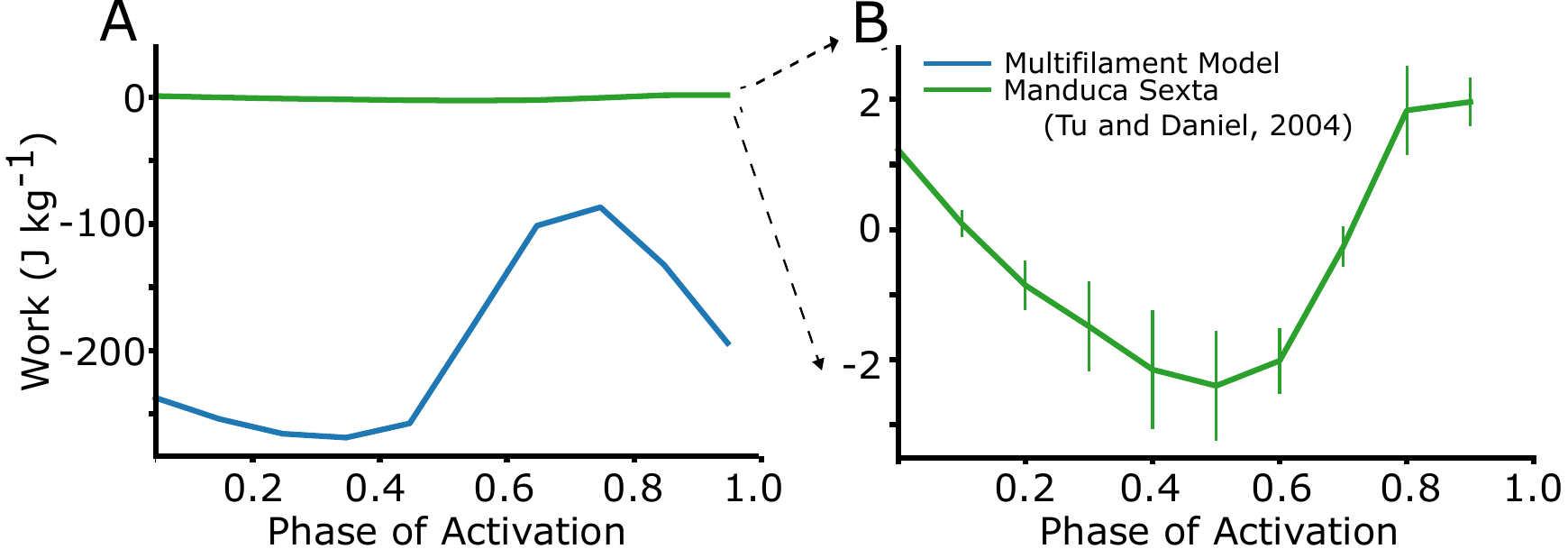}
\caption{\textbf{Work loop simulations based on prior model.} A) Work loop simulations were done at 25 Hz and 10\% peak-to-peak amplitude, which is the \textit{in vivo} frequency and amplitude of \textit{Manduca sexta}. Blue shows the net work at different phases of activation simulated form the previous 2sXB model, compared to green which shows the net work phase sweep for \textit{Manduca sexta} from experiments \cite{Tu_SubMax}. B) The \textit{in vivo M sexta} phase sweep re-plotted to show that net mechanical work changes from positive to negative during the phase sweep, but on a much zoomed in scale. }
\label{fig:w_phase_original}
\end{centering}
\end{figure}

While ideal for capturing axial and radial force contributions, the prior 2sXB models could not produce significant positive work under \textit{in vivo} frequencies and amplitudes. We simulated work loops using the release version of these models at 25 Hz at 10 phases of activation between 0 and 0.9 and compared the results to phase sweep work loop data taken from \textit{Manduca sexta} isolated, whole muscle experiments \cite{Tu_SubMax}. We found that work loops produced orders of magnitude more net negative work (-230 J kg$^-1$ at phase of activation of 0 in simulation compared to ~ 2 J kg$^-1$ in real muscle) under these conditions (Fig. \ref{fig:w_phase_original}). It is important to acknowledge that this dynamic regime with high rates of axial shortening and lengthening were not the purpose of the prior 2sXB model and these simulations only serve to illustrate the regime where modifications are necessary to apply such approaches. Other prior models that did not include a second spring, and hence an explicit radial dependency, could emulate work production under cyclic stress-strain curves, but cannot test the dependency on lattice spacing \cite{Tanner_2008}. Here, we describe the model geometry and adaptations that were made to extend the 2sXB model's dynamic range to \textit{in vivo} strain frequencies and amplitudes.

\subsection{Model geometry}
\begin{figure}[H]
\begin{centering}
\includegraphics[width = 1\linewidth]{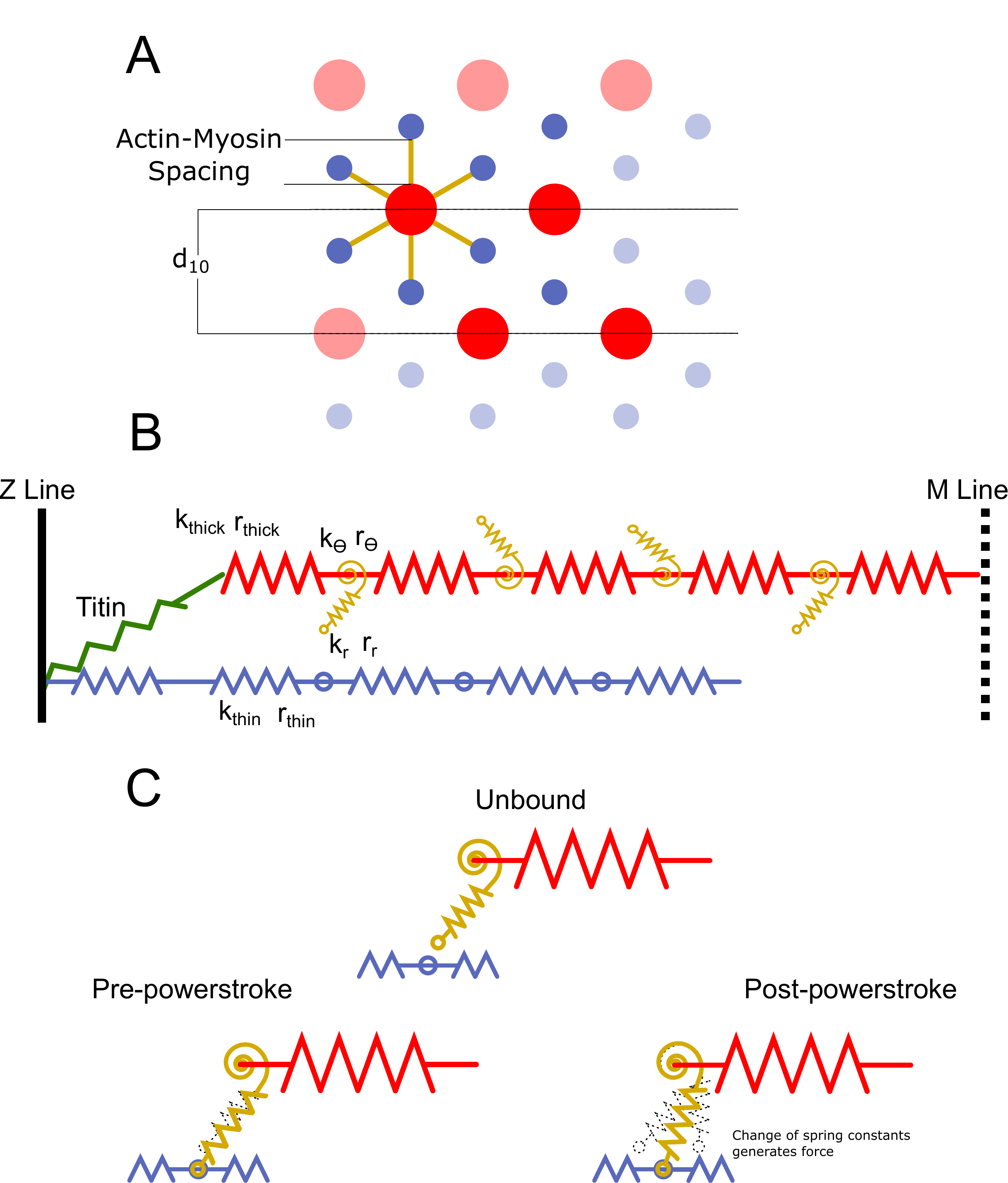}
\caption{\textbf{Half sarcomere geometry and spring element stiffnesses.} The geometry of the spring lattice defines repeating motif that models the half sarcomere. A) A 2-D longitudinal view of a segment of a thick filament and one thin filament with which it interacts. Each myosin head faces a certain actin-containing thin filament with which it can potentially bind. Titin attaches from the end of the thick filament to the z-disk where the thin filament also attaches. B) A cross-sectional view of the half sarcomere, showing the four thick filaments and 8 thin filaments present in model. The $d_{10}$ spacing is the lattice spacing of the crystal unit cell, measured by x-ray diffraction \cite{Irving:2000}. The actin-myosin spacing (minus the diameter of the thick and thin filaments) is the main parameter we vary in the model. C) The thick and thin filaments are composed of series spring elements of stiffness $k_{thick}$ and $k_{thin}$ taken from empirical estimates. Equilibrium lengths are $r_{thick}$ and $r_{thin}$. Each myosin head is governed by a three-state kinetic model, but the free energy of each state is modified by the strain on the head. We use a two spring model for myosin composed of a linear torsional spring at the base ($k_{\theta}$ and $r_{\theta}$) and a linear transitional spring in the arm ($k_{r}$ and $r_{r}$), as in \cite{Williams:2010}. The power stroke is mechanically represented by a change in the rest angle and length of the myosin motor.}
\label{fig:model_description}
\end{centering}
\end{figure}

As in \cite{Williams:2010}, a half sarcomere is represented as a 3 dimensional spring lattice. Myosin-containing thick and actin-containing thin filaments are composed of a series of linear springs (Fig. \ref{fig:model_description}A) where nodes between springs represent either the origin of a myosin motor (in the case of the thick filament) or a potential binding site (in the case of the thin filament). The model consists of 4 thick filaments and 8 thin filaments arranged such that one thin filament is located equidistant between three thick filaments, as in vertebrate muscle \cite{Millman1998}. Each thick filament is attached to the z-disc by titin, which attaches to the z-disk and to the thin filament. This spatially explicit unit (Fig. \ref{fig:model_description}B) is the repeating motif that composes the regular myofilament lattice in a sarcomere. Periodic boundary conditions are enforced so that each thick filament interacts with 6 thin filaments and allow us to scale to arbitrary size. Interactions with the boundary of sarcomere and fluid interaction within the sarcomere are currently ignored.

\begin{table}[ht]
\centering
\caption{Model Spring Parameters}
\begin{tabular}[t]{lcc}
& Pre-powerstroke& Post-powerstroke \\
\midrule
\\
$k_{\Theta}$ & 4000 $pn \cdot nm \cdot rad^{-1}$ & 4000 $pn \cdot nm \cdot rad^{-1}$ \\
$k_{r}$ & 16 $pn \cdot nm^{-1}$ & 16 $pn \cdot nm^{-1}$\\
$r_{\Theta}$ & 47.16$\degree$ & 73.20$\degree$ \\
$r_{r}$ & 19.96 nm & 16.47 nm\\
& & \\ 
& & \\
& & \\
& Thick Filament & Thin-Filament \\
\midrule
$k$ & 2020 $pn \cdot nm^{-1}$ & 14.3 $nm$\\
$r$ & 1760 $pn \cdot nm^{-1}$ & \\
\end{tabular}
\label{table:timing_table}
\end{table}%

Each node of the thick filaments contains triplets of myosin heads, referred to as crowns. The elastic links between adjacent crowns are described as linear springs with a set length of 14.3 nm, consistent with the 14.3 nm repeat in muscle which gives the helical repeat of the myosin heads \cite{Irving_Huxley_M6}. Myosin head triplets are azimuthally distributed by 120\degree{} and adjacent crowns are rotated by 60\degree. Thin filaments are similarly composed of crossbridge binding sites which are spaced 38.7 nm apart and are linked together by linear springs. Filament strain can change the local spacing of heads or binding sites and can arise from either muscle stretch or internal stress produced from myosin binding. The out-of-register nature of myosin heads and binding sites (42.9 nm vs 38.7 nm) is a well-known feature of muscle that emphasizes the importance of a spatially explicit model because compliance in the filaments can either promote or suppress binding probability \cite{Irving:2000, Daniel_1998}. 

The stiffness of the thin filaments $k_{thin}$ were originally estimated in \cite{Kojima_1994_k_thin} from 1 $\mu$m long segments of rabbit skeletal muscle to be 65 pN/nm via deflection of a microneedle under a microscope. The stiffness of the thick filaments comes from the observation that thick filaments are about 150\% stiffer than thin filaments, as seen by strain changes in the thick and thin filaments via x-ray diffraction of frog skeletal muscle \cite{Wakabayashi_1994_k_thick}. The repeat distances of 38.7 and 43 nm are then used to scale the stiffness of each segment of the two filaments \cite{Daniel_1998}.

Myosin heads themselves are deformable and previous spatially explicit models have incorporated either a single linear spring \cite{Tanner_2008, Tanner2007, Pate_cook_1985, Daniel_1998} or two or four springs mixing torsional and linear elements \cite{Williams:2010, Williams:2013bi}. The single linear spring does not accurately capture the radial force that crossbridges generate, or the radial dependence of the binding probability of myosin heads, but the two and four spring models (2sXB and 4sXB, respectively) have given comparable prior results \cite{Williams:2010, Tanner2007}. We therefore use a torsional and linear spring (Fig. \ref{fig:model_description}A). 

Crossbridges are able to bind to and unbind from binding sites on nodes on thin filaments. Myosin binding during muscle contraction has been modeled with many different numbers of states \cite{Lymn1971, Pate_cook_1985, AFHuxley1971}, but based on prior models and because we primarily wanted to look at the effect of myofilament lattice structure on the force production step we focused on a 3-state model where myosin heads can be: 1-unbound, 2-weakly bound, and 3-strongly bound. The transition from weakly bound to strongly bound representing the primary step in force production, called the power stroke (Fig. \ref{fig:model_description}C). The power stroke occurs because ATP hydrolysis causes conformational changes in the structure of the myosin motor, which is represented mechanically as a change in the equilibrium angle and equilibrium length of the torsional and linear springs which comprise the myosin motor \cite{Williams:2010}. The pre- and post-power stroke equilibrium locations of the myosin head come from electron tomography of quick frozen muscle of insect flight muscle \cite{Liu_2006_E_tomography_S2, Taylor_E_tomography_}. 

This model also incorporates titin, a protein filament which attaches the thick filaments to the Z-disk, which defines the end of the sarcomere \cite{JPowers_2018_model}. Each titin filament is connected to each of the four myosin-containing thick filaments at one end, and to the z-disk at the location where the actin-containing thick filament intersects the z-disk. Each titin filament therefore exerts a radial and axial force on the lattice. The force of titin is given by the equation $F_{titin} = a \cdot e^{b \cdot \Delta L}$, as in other models \cite{JPowers_2018_model, Campbell_Ken_2009, Campbell2011a}. For the parameters $a$ and $b$, we used the same parameters as in \cite{JPowers_2018_model}. In real muscle the stiffness of titin is thought to change with $Ca^{2+}$, and is increasingly recognized as an important contributor to muscle function \cite{GRANZIER_2005_titin}, and it has been suggested that titin stiffness could significantly affect work \cite{JPowers_2018_model}. Although titin is present in the model, in the current implementation does not include activation-dependent changes. Furthermore, titin is not present invertebrates like \textit{M. sexta}, although a number of proteins such as sallimus, kettin, and projectin have been identified which may serve an analogous function \cite{YUAN2015, Bullard_sls_2007}.


At the beginning of each time step, transition probabilities are calculated for crossbridge binding and state transitions based on the current state of each myosin head and its distance to the nearest thin filament binding site. The force on each node is calculated as the force from attached crossbridges as well as the force from displaced neighboring nodes. To solve for the equilibrium state of the half sarcomere, each node's axial location is iteratively adjusted so that the instantaneous force on each node is zero. The net force is then calculated as the force exerted by the node nearest the m-line on each thick filament. 

\subsection{Rate functions}

Rate equations for earlier versions of these spatial explicit models were originally established by fitting force under constant velocity data in \cite{Pate_cook_1989} to a model in which crossbridges were represented by linear (axial only) springs. These rates were subsequently adapted in \cite{Daniel_1998, Tanner2007} to include dependence on crossbridge stiffness, and again in \cite{Williams:2010, Williams:2015} to incorporate the radial component of the myosin heads.

The origin of the large negative work in the previous 2xSB models (Fig. \ref{fig:w_phase_original} A) arises from many crossbridges being strained in unphysiological conditions. During a single work cycle at physiological strain velocities, a large population of crossbridges in the prior models transition to the loosely bound state $s_2$ even when strained at ~20 nm, far from their equilibrium strain. They remain attached for some time, being further strained to $\approx$45 nm. This is substantially larger extensions than what a crossbridge should experience, which should be less than 10 nm during rapid shortening \cite{Pate_cook_1985, Pate_cook_1989}. These abnormally strained crossbridges generate large amounts of negative (lengthening) force during shortening. These loosely bound crossbridges are not binding from an unbound state ($s_1$) but rather are reverting from the strongly bound state ($s_3$). This is because the $r_{31}$ rate does not increase rapidly enough at high strains, and reverse power stroke rate $r_{32}$, increases around -20 nm. While this regime of extreme, unphysiological strains were unlikely to have been explored in previous simulations of the 2sXB model that consider isometric conditions, they prevent realistic force under dynamic conditions. 

\begin{figure}[H]
\begin{centering}
\includegraphics[width = 1\linewidth]{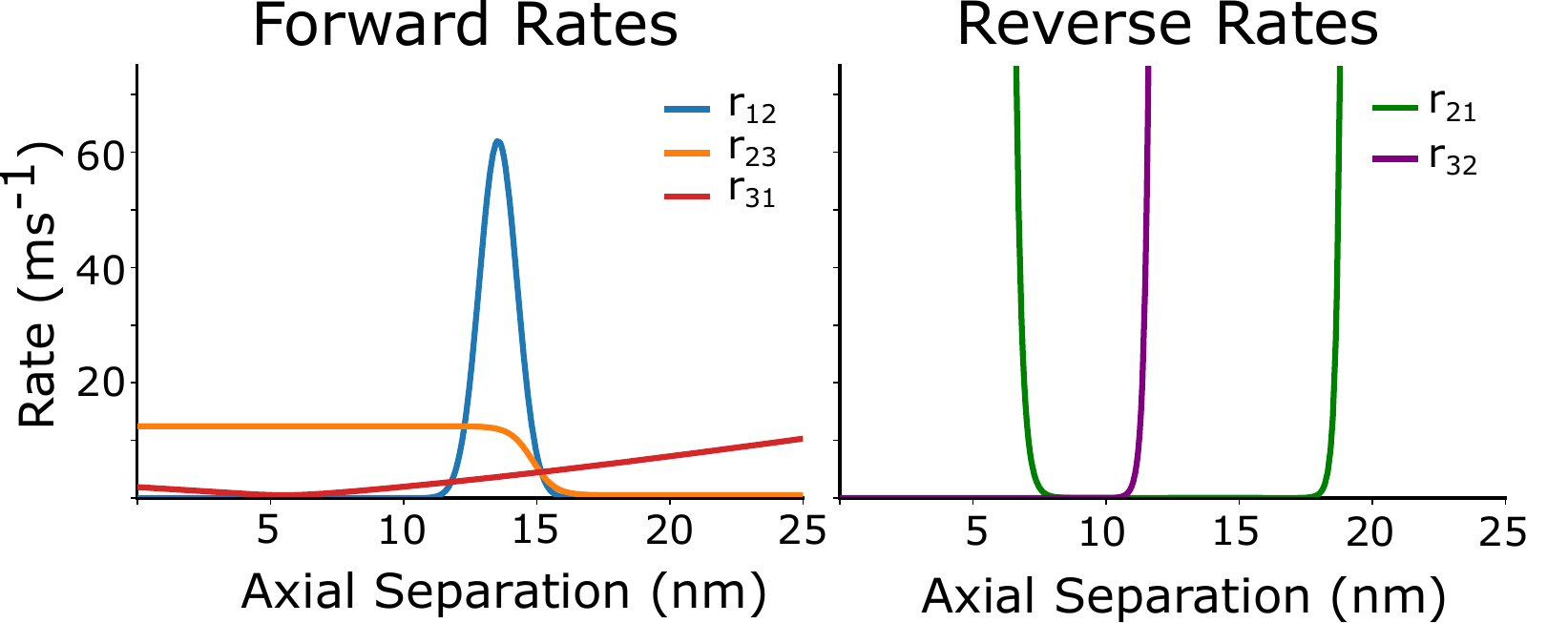}
\caption{\textbf{Rate functions at an actin-myosin spacing of 15, without thermal forcing.} Left) The forward rates $r_{12}$, $r_{23}$ and $r_{31}$ rate. Right) Here we show the reverse rates $r_{21}$ and $r_{32}$. The rate $r_{13}$ is defined to be 0. }
\label{fig:default_rates}
\end{centering}
\end{figure}

The inappropriate reverse transition to $s_2$ and persistence in that state comes from the model exploring the tails of the rate functions. In particular, the unbinding rate $r_{21}$ is the ratio of the binding rate $r_{12}$ and the difference in free energies between two states of the expression $\exp^{U_{1} - U_{2}}$ (Fig. \ref{fig:default_rates}A,B). The falloff of $r_{21}$ is too slow relative to $\exp^{U_{1} - U{2}}$, which causes the unbinding rate $r_{21}$ to be 0 at extreme strains, when it should be rapidly rising. This meant that when tightly bound crossbridges revert from the strongly bound to the loosely bound state, instead of nearly instantly dissociating, they instead became negatively strained up to 40 nm during shortening. Similarly, loosely bound crossbridges would become positively strained during lengthening. The large forces caused by these highly strained crossbridges opposing length change in the sarcomere was the major cause of the negative work being done. 



Because we wanted to maintain consistency with the previous instances of the spatially explicit model as much as possible, we sought to change the behavior of the rate functions by making rates steeper at higher strains without substantially changing their behavior at low strains. Comparing to the rate equations which were originally fit in \cite{Pate_cook_1985, Pate_cook_1989}, we saw that the binding rate $r_{12}$ exponentially decreases with increasing distance from the binding site just as in later versions of the model. However \cite{Pate_cook_1989} also added a baseline rate of .005 ms$^{-1}$ to $r_{12}$ which is not present in the earlier 2xSB models. At first glance this seems nonphysical, since it implies that crossbridges have a chance to bind at any axial distance. However the magnitude is too small to practically change $r_{12}$ significantly, and when we re-examine the $r_{21}$ rate, this baseline offset in $r_{12}$ corrects the problem with the binding rates exponential falloff, which enforces an infinite well in the $r_{21}$ rate without substantially changing binding rates in the working range of the myosin head. The transition rates used here, based on those in \cite{Williams:2010} are given by the following equations: 



\begin{equation*}
\begin{split}
r_{12} &= \tau * e^{-d^{2}} + .005 \\
r_{21} &= \frac{r_{12}}{\exp{(U_{0}-U_{1})}} \\
r_{23} &= A*(1 + tanh(C + D (U_{1} - U_{2}))) \\
r_{32} &= \frac{r_{32}}{\exp{(U_{1}-U_{2})}} \\
r_{31} &= G \sqrt{U_{2}} + H \\
r_{13} &= 0
\end{split}
\end{equation*}


Here, $U_{i}$ is the free energy in the $i^{th}$ state, $d$ is the distance from the myosin head to actin binding site, and the rate constants $\tau$, $A$, $C$, $D$, $G$, and $H$ are chosen so that the function has units of 1 ms$^{-1}$, and the functions yield transitions consistent with previous models \cite{Williams:2010, Williams:2013bi, Daniel_1998, Tanner2007} and experimental data \cite{Pate_cook_1989}. Probability of a transition is calculated from the rate as $1 - e^{-r_{ij} \cdot dt}$, where $dt$ is the time step in the simulation. In our simulations, $\tau$=72, $A$=.8, $C$=6, $D$=.2, $G$=.6, and $H$=.02. Reverse rates are defined by the equilibrium equation $r_{ji} = \frac{r_{ij}}{e^{U_{i} - U_{j}}}$.


While this change was able to account for much of the negative work being done in work loops simulations, we still found that the $r_{21}$ was not tightly constrained compared to previous incarnations of the model \cite{Pate_cook_1989, Tanner2007}, causing crossbridges to become nonphysically strained (Fig. \ref{fig:default_rates}C,D). While individual rate functions could be adjusted, the overall pattern is that myosin heads tend to remain in either $s_2$ or $s_3$ at unreasonably large strains. This is consistent with an underestimation of the effective stiffness of the myosin head. We therefore stiffened the myosin head's torsional spring by a factor of 10 compared to the previous model. This affects the $r_{21}$ rate since it is dependent on the free energy of the myosin head, which is dependent on the stiffness of both spring elements, and also makes the $r_{31}$ rate steeper \cite{Daniel_1998, Tanner2007, Williams:2010}. We chose to increase the torsional spring stiffness since it is the dominant contributor to the steepness of the rate equations in the axial direction. 

After these changes we found that the model produced much less tetanic force than the peak tetanus force of \textit{Manduca sexta} DLM. We also found that the dominant contributor of force was from the loosely bound state, while the tightly bound state contributed little net force. Therefore, we also chose to increase the stiffness of the myosin head's linear spring by a factor of 4, and the power stroke rate constant by a factor of 10. We chose these parameters because this set the average steady-state force of a crossbridge to be about 8-10 pn under isometric tetanus, consistent with estimates of the force of the crossbridge power stroke \cite{Piazzesi_xb_2002}. Besides more closely matching these physiological observables, increased binding might be expected to match to data from invertebrate flight muscle because the original model in \cite{Pate_cook_1989} was derived from rabbit psoas, a slower muscle than \textit{M sexta} flight muscle \cite{Caterini_mouse_WL,Askew_1997_Mouse_WL_fatigue}. Although the stiffness we use is larger than what has been reported from single molecule experiments, these experiments may underestimate stiffness \textit{in vivo} \cite{Piazzesi_xb_2002, Molloy_xb_force_1995, Daniel_1998}. 


\begin{figure}[H]
\begin{centering}
\includegraphics[width = 1\linewidth]{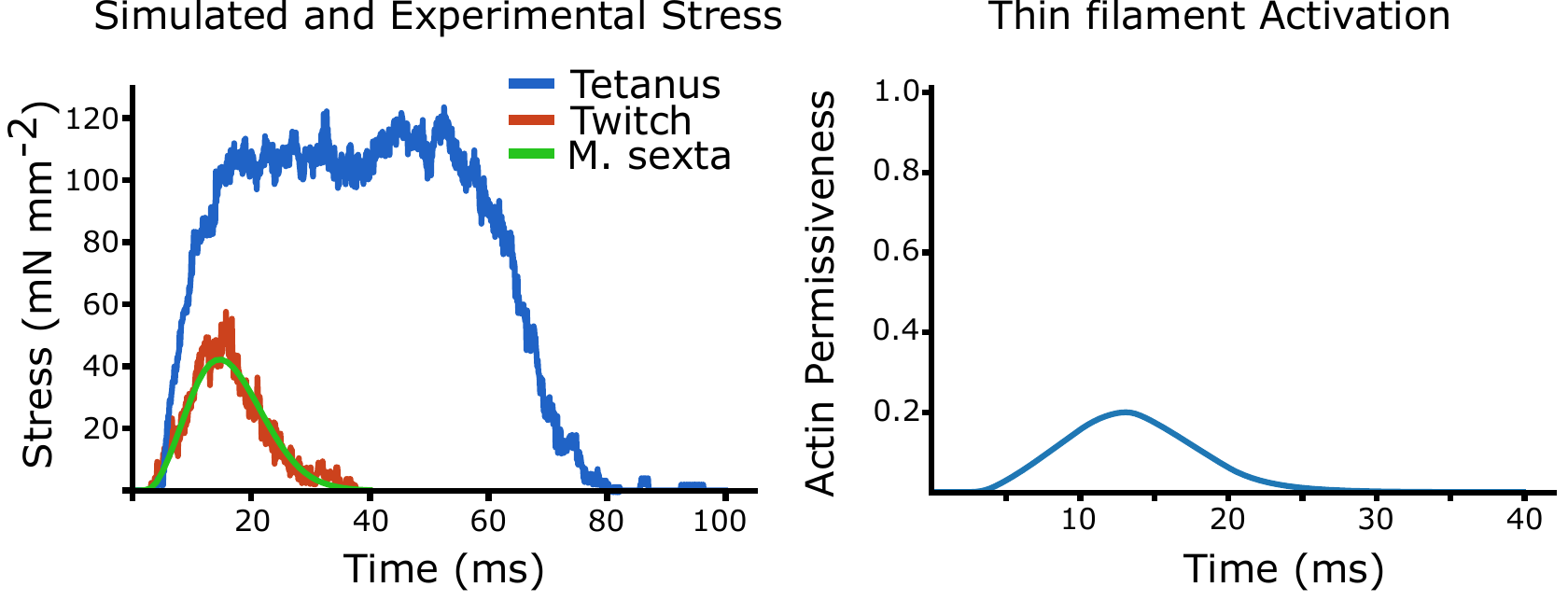}
\caption{\textbf{Simulated twitch and tetanus} We plot the peak isometric tetanic force, then the activation curve which yielded twitch like-force \textit{M. sexta}. This activation curve is then used in all the following work loop simulations. }
\label{fig:twitch_tetanus}
\end{centering}
\end{figure}

\subsection{Actin permissiveness parameterizes Ca$^{2+}$ and tropomyosin dynamics}
In passive real muscle, actin bindings sites are obscured by tropomyosin, which wraps helically around actin and is regulated by the troponin complex of proteins. When a muscle is activated, Ca$^{2+}$ rapidly floods the sarcomere, binds to troponin C, which causes a conformational change in tropomyosin, allowing myosin heads to attach. When Ca$^{2+}$ is pumped out of the contractile lattice, tropomyosin reverts to its original confirmation, preventing myosin binding and force generation. This entire process is parameterized in the model by a single 'actin permissiveness' value which is bounded from 0 to 1 and represents the availability of an actin binding site for potential myosin binding. The product of actin permissiveness and the binding probability calculated from spatial configuration equals the actual probability of binding. An actin permissiveness of 0.5 would indicate that only half of the actin binding sites in the sarcomere are available for binding, which in the model is handled as each instance of binding being 50\% less likely. The actin permissiveness is the same for each binding site in the sarcomere even though the binding probability of a given site will depend both on this and the spatial arrangement of available myosin heads.

\subsection{Activation profile was found by matching to twitch force}
Since work loops are cyclically activated, we needed to define a periodic function for the actin permissiveness, or activation curve, for the sarcomere. We set the shape the actin permissiveness curve as two exponential functions representing influx and re-uptake of Ca$^{2+}$. We then simulated an isometric twitch by choosing the influx time and half life of Ca$^{2+}$ re-uptake such that the rise, fall, and peak force during model response matched twitch data taken from \textit{M. sexta}. The simulated tetanic force and twitch force are shown in figure \ref{fig:twitch_tetanus}, as well as a twitch from \textit{M. sexta}. We used this same activation curve in all following work loop simulations.


\subsection{Actin-Myosin Spacing and $d_{10}$}

We wanted to see if changes in the actin-myosin spacing in a half sarcomere model could modify work output. Since the actin-myosin arrangement in muscle is highly ordered, x-ray diffraction can be used to measure the $d_{10}$ spacing. However, $d_{10}$ is a measurement of the size of the crystallographic unit cell, not a direct measurement of the actin-myosin spacing. It is however, proportional to the actin-myosin spacing, with the proportionality constant depending on the type of muscle. Vertebrate muscle, invertebrate limb muscle, and invertebrate flight muscle all have different proportions and arrangements of actin relative to myosin \cite{Millman1998, Squire_muscle_review, Tune_2020}. In vertebrate muscle, the actin-myosin spacing is given by $\frac{2}{3}d_{10}$ and in invertebrate flight muscle it is $\frac{1}{\sqrt{3}}d_{10}$. Vertebrate $d_{10}$ spacings are typically in the range from 35-40 nm, whereas invertebrate $d_{10}$ spacings tend to be larger, ranging between 40-50 nm. This model and piror 2sXB models, use binding rates originally fit to vertebrate force-velocity curves, and a actin-myosin geometry from vertebrate muscle. We centered our simulations on an actin-myosin spacing of 15 nm (when the radius of the thick and thin filaments are subtracted out), which corresponds to a $d_{10}$ of 47.5 nm in invertebrate muscle, the average value for \textit{M. sexta} \cite{Malingen2020}. Because of the different packing arrangements in vertebrate muscle, even though the $d_{10}$ spacing is different from vertebrate muscle, the actin-myosin spacing is similar (12.8 nm actin-myosin spacing at 38 nm $d_{10}$).

\section{Results}

\subsection{Simulated work-phase sweep captures main features of \textit{M sexta} work-phase relationship}

After tuning our model to twitch and tetanus data from \textit{M sexta}, we first tested if it could capture realistic levels of mechanical work under dynamic, physiological conditions. We simulated these work loops with a peak-to-peak strain amplitude of 10\% and at a frequency of 25 Hz, as in \textit{M sexta} and varied the phase of activation which should cause net work to smoothly transition from positive to negative depending on when myosin heads are actively recruited. Under the crossbridge stiffnesses and rate constant change we made, we simulated work loops at 16 phases of activation. We initially kept a constant lattice spacing of 15 nm, which would correspond to a $d_{10}$ spacing of 47.5 nm in \textit{Manduca sexta}. 

Whereas prior models that incorporate explicit radial strain dependence did not generate any net positive work and were multiple orders of magnitude away from predicting force under dynamic conditions \ref{fig:w_phase_original}, our revised 2sXB model produced a strong match to physiological work loops at all phases. Each trial included 16 periods, and work was calculated for each period and averaged to obtain means and standard deviations. At a phase of activation of 0 - the average \textit{in vivo} phase for flight in \textit{M sexta} - our model produced 0.6 $\pm$ .2 J/kg (mean $\pm$ s.d.), compared to 1.6 $\pm$ .27 J/kg in \textit{M sexta} whereas the earlier 2sXB model predicted -230 J/Kg. At a phase (0.8) that maximized positive mechanical, our updated 2sXB model produced 1.06 $\pm$ .28 J/kg compared to 2.93 $\pm$ .59 J/kg \textit{in vivo}. During phases of activation around the transition from the end of shortening to the beginning of lengthening (0.5), the model produced more negative work than \textit{M sexta}. For example, the model produced -3.5 $\pm$ 0.5 J/Kg, compared to -1.9 $\pm$ .4 J/kg \textit{in vivo} (mean $\pm$ s.d.). Despite not being explicitly tuned to match the dynamic conditions of work loops, the model both captures work output to within a factor of 3 (compared to a factor of $>$100) and shows a phase dependency that matches \textit{in vivo} expectations.

\begin{figure}[H]
\begin{centering}
\includegraphics[width = 1\linewidth]{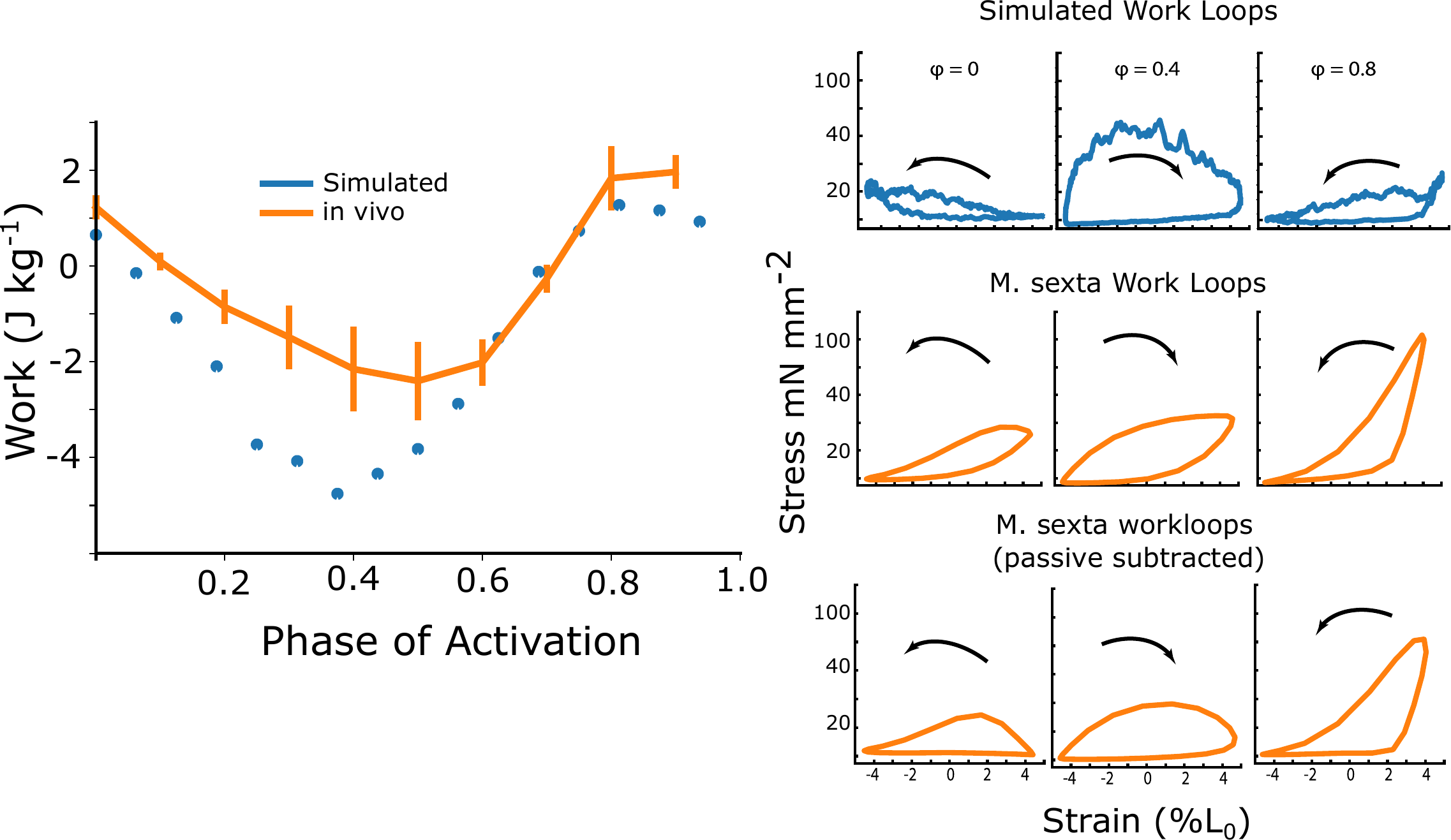}
\caption{\textbf{Simulated net work vs. phase of activation} A) We plot the net work vs. phase of activation produced by our updated 2sXB model (blue) as well as the measured \textit{in vivo} net work for \textit{M sexta} (orange). B) We show example simulated work loops at phases of activation of 0, 0.4, and 0.8. We define $\phi=0$ as the start of shortening. C) We also show real work loops from \textit{M sexta} at the same phases for comparison (Fig. \ref{fig:w_v_phase_base} C). Because the passive component of force in real muscle is much higher than in our model, we show also \textit{M sexta} work loops which have had the passive component of force subtracted.}
\label{fig:w_v_phase_base}
\end{centering}
\end{figure}

Comparing the simulated work loops with real work loops from \textit{M sexta}, there are several notable differences. First of all, there is a large passive component of force in real muscle which is not present in the model. This can be seen from the ramp in force as muscle strain increases (Fig. \ref{fig:w_v_phase_base} C). Because the passive component of force in real muscle is much higher than in our model, we show also \textit{M sexta} work loops which have had the passive component of force subtracted (\ref{fig:w_v_phase_base} D). We found at the \textit{in vivo} phase of activation for \textit{M sexta} of 0 (the start of shortening), that peak passive-subtracted force occurred 5 ms after activation occurred, whereas in simulated work loops the force rose much slower, only peaking 20 ms after activation. At a phase of activation of 0.4 (just before the transition from shortening to lengthening), the force in the simulated work loops rises much faster and higher during the first few milliseconds than in the passive subtracted, however they both exhibit the same plateau of force during lengthening. At a phase of 0.8, the force in \textit{M sexta} work loops is considerably higher than that of simulated work loops, with \textit{M sexta} work loops producing 100 mN/mm$^2$ compared to a peak force of only 40 mN/mm$^2$ in simulated work loops.



Many of these differences likely arise from not specifically matching the model to replicate \textit{M. sexta} parameters. One possible avenue of for future research would be to examine if species-specific structural differences could give tighter fits to specific datasets. For example, we should expect variation in the actin:myosin ratio, the orientation of the repeating lattice unit, and the presence of other active filaments and regulatory proteins influence force production under dynamic conditions. Notably, the passive stiffness of titin has been shown to influence the amount of crossbridge binding and force in a spatially explicit muscle model under isometric contractions at high strain \cite{JPowers_2018_model}. Since the passive component of our model is so low, increasing the stiffness of titin a significant amount could have a large impact on mechanical work. While elaborations could be made to make the updated 2sXB model more like other specific systems, the fundamental formulation here is sufficient to test if structural variation can drive large changes in work output under physiological conditions. Our goal was not to optimally reconstruct work done by a specific insect muscle in a specific context, but rather to obtain a model that has reasonable behavior of insect skeletal muscle under dynamic, oscillatory conditions and then interrogate if lattice spacing can modulate this work in a significant way.


\subsection{1 nm spacing changes can generate positive or negative net work}

\begin{figure}[H]
\begin{centering}
\includegraphics[width = 1\linewidth]{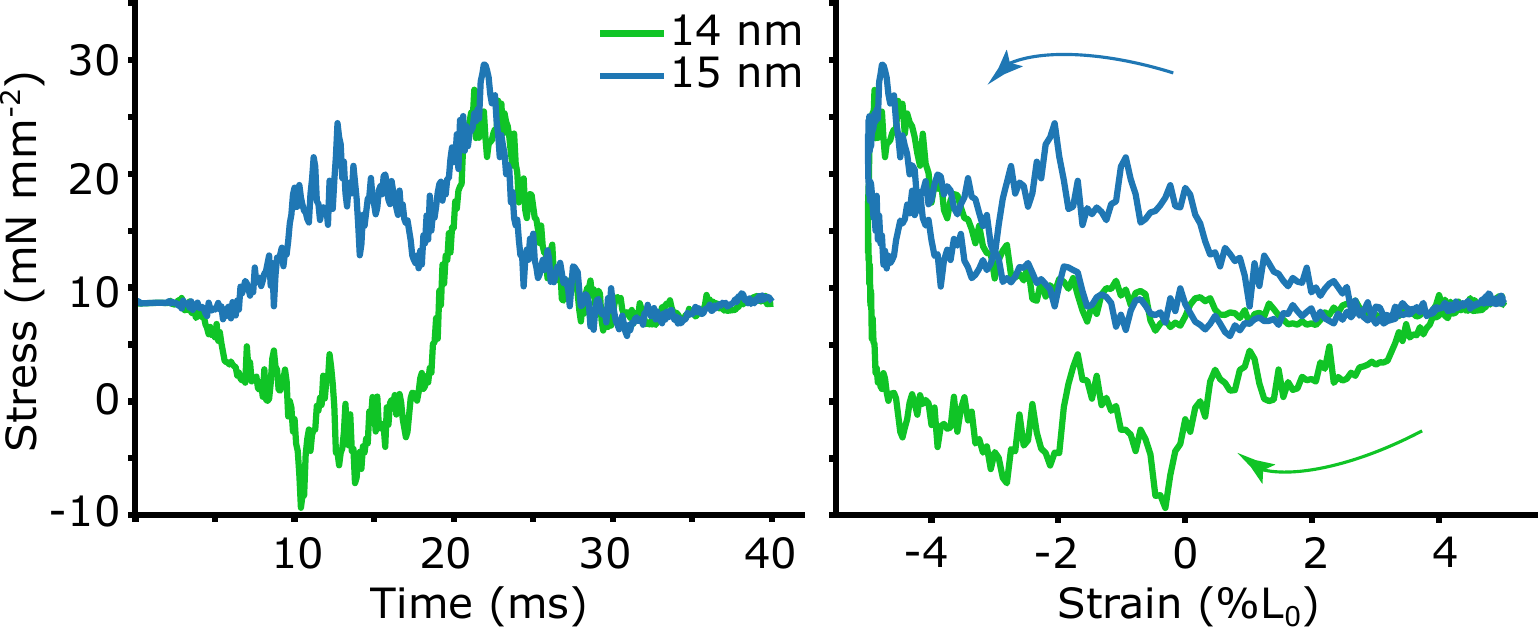}
\caption{\textbf{Brake-like (net-negative) and motor-like (net-positive) work loops with 1 nm lattice spacing change.} We show stress vs time and stress vs strain (work loop) simulated at an actin-myosin spacing of 14 (green) and 15 (blue). Since the model is spatially explicit, there is a high degree of stochasticity, which is why we simulated 16 periods and averaged. }
\label{fig:1_nm_work_}
\end{centering}
\end{figure}

The updated 2sXB model allows us to now test if small differences in the axial spacing of the myofilament lattice can modulate muscle mechanical work, as suggested in \cite{Tune_2020}. After getting a reasonable phase sweep at 15 nm, we simulated work loops at 14 nm. In invertebrate flight muscle, this would correspond to a $d_{10}$ change of 47.6 to 45.9, a 1.7 nm difference. We found that under these conditions, at a lattice spacing of 14 nm the net work was negative (-0.74 $\pm$ 0.14 J/kg), while at 15 nm, the model produced net positive work (0.72 $\pm$ 0.14 J/kg) (Fig. \ref{fig:1_nm_work_}). A single nanometer difference in lattice space can cause a switch in the sign of the model's work output.

We next extended the simulation to lattice spacings from 12 to 17.5 nm, again keeping lattice spacing constant throughout the entire work loop. At the \textit{in vivo} phase of activation lattice spacing had a net work peak at 16 nm (Fig. \ref{fig:W_ls_isoV_isoLS}, $\phi=0.0$, red). As lattice spacing increased from 12 to 16 nm, net work changed from -4.2 J/Kg to 1.3 J/kg, increasing positive work by 1.3 J/kg nm$^{-1}$. Similarly, at a phase of activation of 0.85, the net work peaked at a lattice spacing of 15.75, with net work increasing 3.0 J/kg nm$^{-1}$ from 12 to 16 nm. In contrast at a phase of activation of 0.15, net work only slowly increases with lattice spacing, and never peaks. The peak in the phase of activation occurs at an actin myosin spacing equivalent to a $d_{10}$ of 49 nm, while the recorded mean $d_{10}$ spacing in \textit{M sexta} is 47 nm \cite{Malingen2020}.


\subsection{Isovolumetric and \textit{in vivo} lattice spacing dynamics increased net work by 10-20\%} 

\begin{figure}[H]
\begin{centering}
\makebox[\textwidth][r]{\includegraphics[width=1\textwidth]{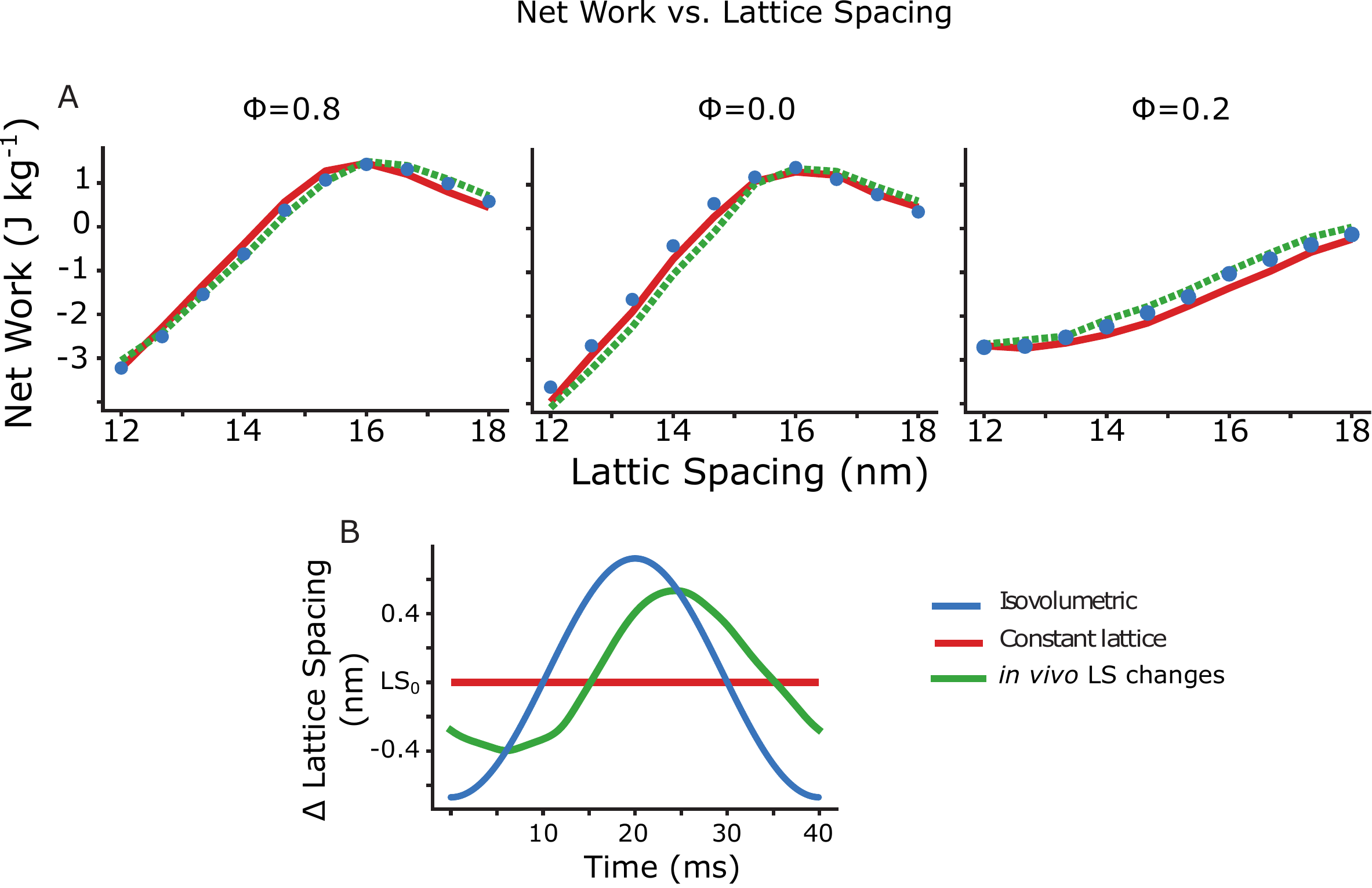}}
\caption{\textbf{Net work vs. lattice spacing under isovolumetric lattice changes, constant lattice, and \textit{in vivo} lattice changes from \textit{M. sexta}} A) We show the net work at phases of activation of 0.8, 0.0, and 0.2 under conditions in which the lattice spacing was either constant (isolattice, red), changed with sarcomere strain with a Poisson's Ratio ratio of 0.5 (isovolumetric, blue), or was prescribed according to the \textit{in vivo} changes found from \textit{M. sexta} work loops at a phase of activation of 0. B) shows the prescribed lattice spacing changes in the different cases centered on L${_0}$.}
\label{fig:W_ls_isoV_isoLS}
\end{centering}
\end{figure}

We next wanted to show how the net work would depend not only on the mean offset of the lattice spacing, but on the amplitude of the spacing change. In many muscles the lattice spacing in not constant, but depends on the length of the sarcomere \cite{Dickinson:2000vg, Malingen2020, Irving:2000}. We wanted to test how work would be influenced when we made the actin-myosin spacing depend on sarcomere length. To start with we chose to make the lattice spacing isovolumetric with length change. We then prescribed a time course of lattice spacing change similar to what has been recorded in \textit{M. sexta} by x-ray diffraction during work loops. We then compared the net work under these different conditions- constant lattice spacing (termed isolattice), constant volume (isovolumetric), and \textit{in vivo} lattice spacing changes. Fig. \ref{fig:W_ls_isoV_isoLS} B shows the time course of lattice spacing changes for the different conditions, where LS$_{0}$ is the lattice spacing at the mean strain. 

We simulated work loops at 25 Hz with the same activation and strain pattern used in work loops as above, and 10\% peak-to-peak strain amplitude. Each point in figure \ref{fig:W_ls_isoV_isoLS} is the average of 20 periods of cyclical activation and strain. Isovolumetric conditions indicate the lattice spacing changed with length according to the equation $ \Delta d = d (1-(1+\frac{\Delta L }{L})^{-\nu})$, where $d$ is the $d_{10}$ spacing, which we then convert to face-to face actin-myosin spacing, and $L$ is the length of the simulated half sarcomere. The Poisson ratio is given by $\nu$, and $\nu=.5$ will give isovolumetric changes. The \textit{in vivo} lattice spacing changes have a time-varying Poisson ratio. We simulated here three phases of activation. A phase of 0 is the \textit{in vivo} phase, with $\phi$=0.85 and $\phi$=0.15 being the limits of the \textit{in vivo} range in \textit{M sexta}. 

We found that while the mean spacing was a much more dominant factor in determining net work overall, the time course of the lattice spacing change could still have a small affect on net work. For example at a phase of activation of 0.2, activation begins at 8 ms and peak actin permissiveness occurs 13 ms later (at 21 ms). Maximal activation therefore coincides with the peak lattice spacing change in the isovolumetric and \textit{in vivo} cases. This increases net work by a small amount. In the case of isovolumetric change the work enhancement over constant lattice conditions is approximately constant between lattice spacings of 14 to 18 nm and increases by an average of 0.22 J/kg. Under \textit{in vivo} lattice spacing changes, the net work enhancement is larger (average of 0.35 J/kg over 14-18 nm), even though the peak lattice spacing change is smaller, possibly because the lattice spacing change is phase advanced compared to isovolumetric changes. This would allow for a larger mean lattice spacing for the portion of the work loop following peak activation. At a lattice spacing of 15, this represents an increase in positive work of 12\% for isovolumetric and 21\% for \textit{in vivo} lattice spacing changes compared to constant lattice. In contrast, the net work changes during a phase of activation of 0.8 are minimal for the three cases, since peak activation would occur at around 5 ms, when differences in lattice spacing between the three conditions changes are smaller. Had the phase shift between the \textit{in vivo} lattice spacing changes been larger, we might have seen larger dependencies on time course of lattice spacing change suggesting that lattice spacing dynamics may have larger effects during large strain and high-frequency behaviors. 






\subsection{Cross bridge stiffness can attenuate or accentuate net work dependence on lattice spacing}

\begin{figure}[H]
\begin{centering}
\includegraphics[width = 1.\linewidth]{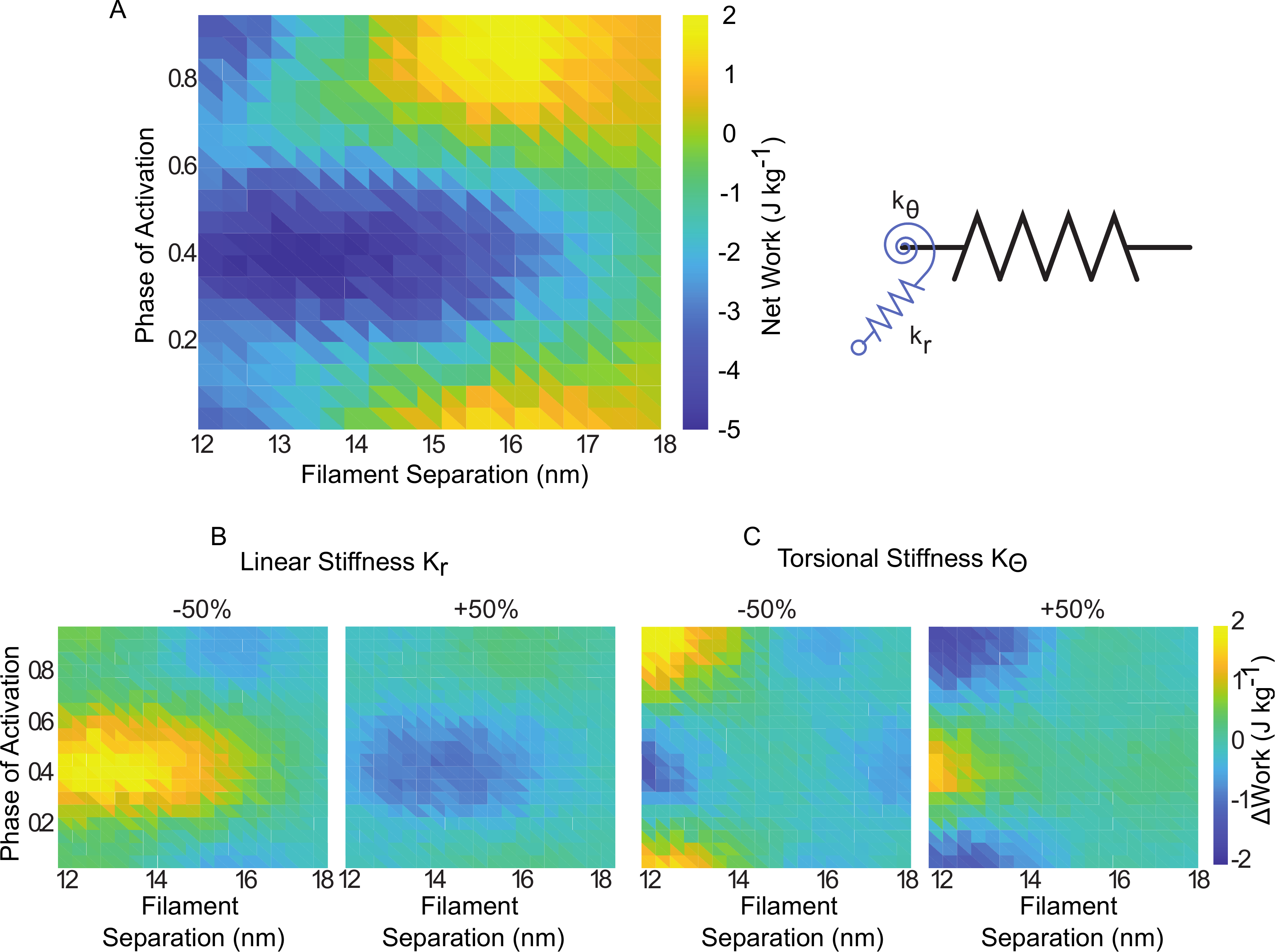}
\caption{\textbf{Net work vs phase of activation and lattice spacing.} A) We simulated work loops in the half sarcomere model at phases of activation of 0 to 0.95 in 0.05 increments, as well as over lattice spacings from 12 to 18 nm. B) We then simulated work loops over the same range, but with the stiffness of the linear and torsional springs increased or decreased by 50\% separately. Data in B) and C) are shown as change relative to A. }
\label{fig:deafault_stiff_surf}
\end{centering}
\end{figure}

Because we expect the stiffness of the springs composing the myosin heads to affect net work, we next changed the stiffness of the linear and torsional springs and calculated the amount of work generated by the half sarcomere simulation (Fig. \ref{fig:deafault_stiff_surf} A). We found that in the default state of the model ($k_r$=16 pn $\cdot$ nm$^{-1}$, $k_{\theta}$=4000 pn $\cdot$ nm $\cdot$ rad$^{-1}$), the region of greatest net work was at lattice spacings between 15 and 16 nm and phases of activation between 0.8 and 0.1 (end of lengthening to mid-way through shortening) and the region of minimum net work was between lattice spacings of 12-15 nm, and phases of activation of 0.3 to 0.5 (midway through shortening to start of lengthening). 

We next simulated the same range of lattice spacings and phases of activation, but set the stiffness of the linear and torsional components to $\pm$50\% the default values (Fig. \ref{fig:deafault_stiff_surf} B, C). We found that changing the linear stiffness changed the net work for phases of activation near the start of lengthening ($\phi$=0.5), and that torsional stiffness mostly affected phases near the start of shortening ($\phi$=0). In both cases decreasing the stiffness led to an increase in net work in the areas where negative work was largest, meaning they would be less capable of dissipative behavior. Increasing the stiffness had the opposite effect, allowing the linear and torsional spring to have a larger dissipative (negative) work, making them more brake-like.


\subsection{Titin exponential stiffness changes did not affect net work}

\begin{figure}[H]
\begin{centering}
\includegraphics[width = .75\linewidth]{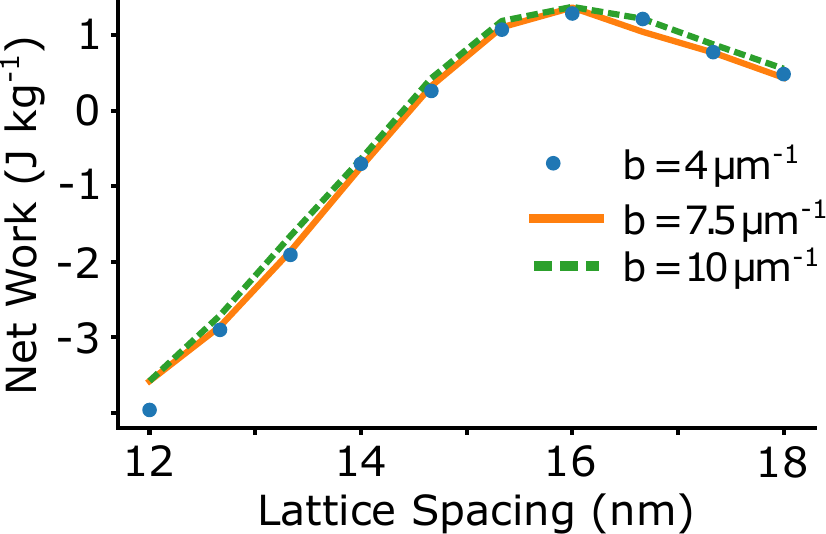}
\caption{\textbf{Net work vs phase of activation with different exponential stiffnesses of titin.} A) We simulated work loops in the half sarcomere model at phases of activation of 0 to 0.95 in 0.05 increments, with the same activation profile and crossbridge stiffness as in Fig. \ref{fig:w_v_phase_base}. The condition $b=4 \mu m^{-1}$ is the condition used in all simulations above. We found that the titin stiffness did not effect the net work under the conditions we examined.}
\label{fig:titin_phase_vs_ls_work}
\end{centering}
\end{figure}

Because titin and titin-analogs in muscle might regulate lattice spacing dynamics \cite{Fukuda:2003, Fuchs2005}, we also simulated half sarcomere work loops under varying exponential stiffness. The force of titin is here modeled as $F_{titin} = a \cdot e^{b \cdot \Delta L}$, and we set $a = 260$ pN and we varied the parameter $b$ from 4 - 10 $\mu m^{-1}$, as in \cite{JPowers_2018_model}, which covers the reported range of estimates for single titin molecules \cite{Linke_1998_titin_elasticity}. It was previously seen in \cite{JPowers_2018_model} that isometric force was diminished when the half sarcomere was at lengths greater than 2.7 $\mu$m. In contrast, we did not find that increasing the stiffness of titin had a large impact on the net work, likely because we did not investigate the same sarcomere length range. Since we based our work loop simulations on \textit{M sexta} DLM, we chose a 10\% peak-to-peak amplitude around a sarcomere length of 2.5 $\mu$m, which meant we did not examine the regime with sarcomere lengths large enough to cause reduced force in the force length curve \cite{JPowers_2018_model}. The effect of titin stiffness on work depends on what regime the muscle is operating in.

\section{Discussion}

The updated 2sXB spatial explicit model can simulate realistic scales of mechanical work under dynamic conditions and supports the hypothesis that nanometer scale changes in the myofilament lattice can significantly effect the mechanical output of whole muscle. Previously, it was shown that lattice spacing differences on the order of 1 nm in two muscles in the cockroach \textit{Blaberus discoidalis} were associated with their different mechanical functions \cite{Tune_2020}. However, it could not be definitively shown that the lattice spacing differences observed were responsible for, rather than just correlated with modulating work. Our updated 2sXB model results demonstrate that differences in mean lattice spacing alone, even at the scale of a single nanometer, can drive mechanical functional differences, for example switching a motor to a brake. 

\subsection{Lattice spacing and crossbridge stiffness mediate multiscale interactions that alter whole muscle function}

Even though the amplitude of lattice spacing change over the course of one contraction cycle is only a few nanometers, it can have a large effect on force production because it affects the binding rates of all of the billions of myosin motors. Under most physiological conditions, binding transitions are less likely at higher lattice spacings, as are power stroke transitions. However binding events are also able to generate more force, enough to cause positive work, in contrast to the net work at lower lattice spacings (Fig. \ref{fig:W_ls_isoV_isoLS}). Furthermore, this effect can be multiscale, since the lattice spacing of a sarcomere will depend on the length of each sarcomere, which may not be uniform in the whole muscle \cite{Julian_FL_sarc_hetero_1979, Campbell_Ken_2009}. This means that as muscle oscillates, the lattice spacing can be a determinant of muscle force. This force strains the filaments within each sarcomere altering length and lattice space, which in turn again affect force \cite{Campbell2011a, Campbell2011b}. 

There are two sources of lattice spacing changes to consider - time-varying changes and static lattice spacing offsets. Time varying lattice spacing changes can arise from a combination of passive axial strain and the active production of radial force by myosin crossbridges. Static offsets in lattice spacing between two muscles, as seen in the cockroach muscles \cite{Tune_2020}, could arise from different radial stiffness and equilibrium position in the lattice likely determined by the titin-like anchoring proteins and the structure of the z-disk. While we do not know how the lattice spacing's relationship with length is set in muscle, it seems to be muscle specific \cite{Irving_Maughan:2000, Malingen2020, Tune_2020}. While we saw a large effect due to changes in lattice spacing offsets \ref{fig:w_v_phase_base}, we only saw a small change due to the time course of lattice spacing change \ref{fig:W_ls_isoV_isoLS}, possibly since the amplitude of lattice spacing change is small. However, even though real muscle has a very complicated structure, including many more elements than are in our model, we are still able to show the potential for lattice spacing to affect net work. 

These effects provide plausible explanations for the functional differences observed in physiological studies of whole muscle. The difference in work between the pair of muscles in \textit{Blaberus discoidalis} was 2.386 $\pm$ 1.8 \cite{Ahn:2006dy}, and the $d_{10}$ spacing difference was 1 nm \cite{Tune_2020}. In our model with a 1 nm actin-myosin spacing change, meaning a 1.8 nm $d_{10}$ change, the model produced a difference of 0.6 J/kg. While our simulated work differences are not as large as those in \textit{Blaberus discoidalis}, it was not precisely tuned to these particular muscles. The most important thing to recognize is that while we were motivated by results from the cockroach \textit{B discoidalis}, we chose to ground our model's behavior in twitch, tetanus, and work loop data taken from \textit{M sexta}, due to the availability of the necessary physiological measurements, biophysical parameters, and the very detailed time-resolved measurements of the lattice spacing of \textit{M sexta} during work loops of varying phases of activation \cite{Cass_sarc_breathing}. Because of this, work loops were simulated at the frequency and amplitude of \textit{M sexta} flight muscle rather than \textit{B discoidalis}. In general, our goal was not to model one specific kind of muscle, but rather to show that lattice spacing offsets and dynamics can affect muscle mechanical function. However, the generalizations we had to make in the model, we found that we were able to produce physiologically plausible levels of force and work under conditions similar to that of \textit{M sexta} and that the lattice spacing could influence the muscle mechanical function. 

In addition to lattice spacing, the stiffness of the filaments and crossbridges has been shown to be a key parameter in muscle force production in prior spatially explicit models \cite{Daniel_1998, JPowers_2018_model, Williams:2010, Tanner2007}. In general, there is a trade-off in that high compliance in the thick and thin filaments allow more crossbridge binding, but less force per crossbridge \cite{Tanner2007}. Also, by increasing the stiffness of the myosin heads, thermal forcing in the unbound state is reduced, which can reduce the effective distance at which heads can bind. Higher stiffness, however, can increase the force that each crossbridge can produce. By altering the crossbridge stiffness in conjunction with the lattice spacing and phase of activation, we were able to test how crossbridge stiffness attenuated or accentuated the work landscape. Changes to stiffness had a much larger impact when the thick filament was closer to the thin filament, and the torsional spring mostly affected net work during active shortening, while the linear spring affected work during active lengthening \ref{fig:deafault_stiff_surf}. These kinds of interacting effects currently can only be shown in a spatially explicit model, where the geometry as well as the biophysics of present elements can be investigated. 

\subsection{3D spatially-explicit models enable interpretation of muscle's multiscale effects}

By changing the behavior of the rate strains in relation to high frequency, large amplitude strain changes, we are able to simulate work loops in a physiological regime. We were then able to show that modulating the lattice spacing can change net work during high frequency work loops with periodic activation. Spatially explicit models of muscle allow for studying how the geometry and mechanical coupling of the myosin motors can impact force and work while incorporating interaction due to deformation of the myofilament lattice. Even when the myosin motors themselves remain unchanged, effective changes in their dynamics can occur due to multiscale interactions, for example, enhancing crossbridge binding by altering filament stiffness alone \cite{Daniel_1998}. These kinds of models capture dynamics that mass action models alone are not able to account for these kinds of multiscale, emergent behaviors. While spatially explicit models can be more computationally intensive, machine learning methods can be used to develop emulators \cite{Kasim_2022_emulator}. These emulators mimic the original model while being much faster and will catalyze broader use of these models in the future.


Even with the refinements here, the 2sXB spatially explicit model does not contain all possible factors contributing to muscle force. While sufficient to test general dependencies like the sensitivity of mechanical work at the macroscopic scale on nanometer scale lattice spacing, further refinements may enable these models to better match specific muscle conditions. For example, the effects of activatable titin could have a large effect on the amount of work produced. In \cite{JPowers_2018_model} it was shown that by increasing the exponential stiffness of titin, crossbridge binding could be increased at high strains, however force of each crossbridge was lower. They also predicted that stiffening titin could also decrease the negative work produced. We were unable to see a difference in the net work produced in the model under the same stiffness values. However, titin is not simply a passive exponential spring, but may have Ca$^{2+}$ dependent properties \cite{GRANZIER_2005_titin, Kiisa_titin_2018}. By introducing activatable titin, for example by making the stiffness of titin dependent not only on length but also the actin permissiveness, we might expect an even more dramatic dependence of net work on lattice spacing.

Another possible improvement is to make the muscle dynamics in the radial axes. Crossbridge attachment in muscle generates not only an axial force, but a radial one as well \cite{Williams:2010, Cecchi:1990}. This radial force can either push or pull the actin-myosin spacing in or out. This means there is a coupling between the lattice spacing and crossbridge recruitment. Currently we prescribe the lattice spacing changes, meaning there is no explicit coupling between radial spacing and radial force of the crossbridges present in the model currently. However, future versions could incorporate this behavior, after measuring the effective radial stiffness of the lattice. This could affect the work since crossbridge recruitment could deform the lattice, resulting in more or less crossbridge recruitment. This would require experiments aimed at measuring the radial stiffness of the lattice and sarcomere. 


However, even it is current form the model presents an opportunity to study how the geometry of other features of muscle structure affect muscle mechanical function. For example, it has been recognized that the thin-thick filament ratio and arrangement in different muscles can be very different. Vertebrate muscle has a 2:1 thin:thick filament ratio, invertebrate flight muscle has a 3:1 thin:thick filament ratio, and invertebrate limb muscle has a 6:1 thin:thick filament ratio \cite{Millman1998}. While the specialization seen in various kinds of invertebrate muscle might be indicative of some functional consequence for these thin:thick filament packing patterns, it has not been investigated what this might be. Isolating the effect of different geometries in sarcomere structure would be very experimentally difficult, whereas in a spatially explicit model the geometry of crossbridge motors and actin binding sites can be examined.




\section{Conclusion}

We were able to show in a spatially explicit model with prescribed radial spacing differences that we could obtain physiological amounts of force and net work. We showed that the lattice spacing could affect the net work in such a model. This model provides a framework for examining how the biophysics and geometric arrangement of force producing crossbridges in muscle can scale through sarcomeres to the whole muscle scale.



\section{Model availability}

This model is available at \url{https://github.com/travistune3/multifil_titin/tree/multifil_manduca_workloops}.

\section{Acknowledgments}

We would like to thank Tom Daniel, Dave Williams, Joe Powers, and Anthony Ascencio for their helpful discussions.

This work was supported by grant W911NF-14-1-0396 from the Army Research Office, National Science Foundation Early Career Development Award MPS/PoLS 1554790, and the Georgia Tech Dunn Family Professorship to S.S. Other support was provided by the National Science Foundation SAVI student research network in physics of living systems (1205878). The original 2sXB model was developed with additional support from Army Research office grant W911NF-13-1-0435. This research used resources of the University of Washington Center for Translational Muscle Research, supported by the NIH National Institute of Arthritis and Musculoskeletal and Skin Diseases under Award Number P30AR074990. This work was also supported by NIH grants R01HL157169, R01HL142624 and an American Heart Association Collaborative Sciences Award.

\nolinenumbers

\bibliography{00ploslatextemplate.bib}

\end{document}